\documentclass[runningheads]{llncs}
\usepackage{graphicx}
\usepackage{tikz}
\usepackage{siunitx}
\usepackage{amsmath}
\usepackage{pgfplotstable}
\usepackage{pgfplots}
\usepackage{filecontents}
\usepackage{tabularx}
\usepackage{longtable}
\begin{document}
\title{Underwater Image Enhancement Using Pre-trained Transformer}
%
%
\author{Abderrahmene Boudiaf\inst{1} \and
Yuhang Guo\inst{1} \and
Adarsh Ghimire \inst{1} \and
Naoufel Werghi\inst{1} \and
Giulia De Masi\inst{1,2}\orcidID{0000-0003-3284-880X} \and
Sajid Javed\inst{1} \and
Jorge Dias\inst{1}}
\authorrunning{A. Boudiaf et al.}
%
\institute{Khalifa University, Abu Dhabi UAE \and
ARRC, Technology Innovation Institute, Abu Dhabi UAE
}
\maketitle              

\begin{abstract}

The goal of this work is to apply a denoising image transformer to remove the distortion from underwater images and compare it with other similar approaches. Automatic restoration of underwater images plays an important role since it allows to increase the quality of the images,  without the need for more expensive equipment.
This is a critical example of the important role of the machine learning algorithms to support marine exploration and monitoring, reducing the need for human intervention like the manual processing of the images, thus saving time, effort, and cost. This paper is the first application of the image transformer-based approach called "Pre-Trained Image Processing Transformer" to underwater images. This approach is tested on the UFO-120 dataset, containing 1500 images with the corresponding clean images.

\keywords{Vision transformer  \and Underwater imaging \and Image enhancement.}
\end{abstract}
 
\section{Introduction}

The ocean covers more than the $70\%$ of our planet, being a large source of energy, food and materials. Yet, it is still in large part unknown, given the technological challenges that have to be faced during any underwater mission: poor visibility, high hydrostatic pressure, poor communications (essentially still based on acoustic systems), absence of GPS with consequent challenging localizaton and navigation.  Despite these challenges, many kinds of underwater missions are necessary to preserve the environment, explore deep sea and support human activities. Scientific missions are oriented to exploration and monitoring of underwater geology and ecosystems; other missions aim to find and explore submerged  archeological sites (like sheep wrecks along commercial sea routes from ancient times, that can improve the knowledge of past civilations). 
Technological missions are oriented to exploitation of underwater resources (oil, natural gas, minerals). Nowadays, offshore wind farms are becoming more and more common, providing a source of clean energy. Fish farms installations are also increasing, to provide additional source of food; this phenomenon is also expected to increase, given the predicted worldwide shortage of food in the coming decades. Finally, rescue missions are very relevant, given the intense marine traffic for transportation of goods and people that occasionally may lead to accidents, often causing natural disasters and risk of loss of life of men overboard. All these activities become even more challenging in harsh environments, like very deep ocean or polar areas. 

For all the above reasons, underwater data, images and videos collection and processing become more and more important to support marine operations at sea. Underwater imaging is particularly challenging because the images suffer from  light absorption, light backscattering due to suspended particles, color attenuation that drastically affect the quality of the image, especially at greater depths (where the natural light is poor), or in dirty water. Colours and contours are both lessen in the underwater environment producing a  blurry image with indistinct  background, where the blue color is dominating as it can be easily explained based on the optical spectrum of the water. These are specific drawbacks affecting the performance of operative tasks like   image-based underwater navigation algorithms, object detection or fish tracking. On the other hand, due to the limited and low-bandwidth communications channels that are nowadays available, the image processing has to be performed on board of underwater robots, enhancing the need for fast and efficient algorithms. 

Image processing has seen rapid improvement since the use of Convolutional Neural Networks (CNNs) for many tasks such as super-resolution, image denoising \cite{TIAN2020117_denosing}, dehazing \cite{cai2016dehazenet_dehazing}, deraining \cite{deraining35} and other image enhancement tasks.  Also in the marine context, many CNN-based networks have been developed to enhance the quality of underwater images \cite{DCP,IBLA,WaterNet,U_GAN,FUnIE-GAN,ufo120}. Meanwhile, image transformer networks have proven to be a strong competitor to the ordinary CNNs in many tasks as shown in a recent work \cite{ipt}. The pre-trained image transformer \cite{ipt} network has been trained to perform four main tasks: denoising, dehazing, and two levels of up-scaling. The results of the Image Processing Transformer (IPT) network showed that the trained IPT in most cases has outperformed the other specialized CNNs in the mentioned tasks. Transformer networks \cite{Image_Transformer,Attention_Is_All_You_Need} usually deal with sequential data. The input sequence consists of a number of indexed units. The sequence is fed to the network which applies attention on the input. The attention operation is done by calculating the product between each two units in the sequence. This results in an exponentially growing number of computations as the length of the input sequence increases. For instance, increasing the the resolution of the image from 100*100 to 200*200 increases the number of pixels by a factor of 4 [\(100^{2}=10,000 , 200^{2}=40,000\)], while the number of computation required by the attention network will increase by a factor of  16 [\((100^{2})^{2}=\num{100e6} , (200^{2})^{2}=\num{1600e6}\)]. Due to the required large number of operations when dealing with images, local attention is used where the image is divided into clusters and the attention is calculated within the cluster \cite{an_image}. The process of relating each pixel with other pixels showed high performance in many tasks that vary from natural language processing to image processing. 

\begin{figure}
\includegraphics[width=\textwidth]{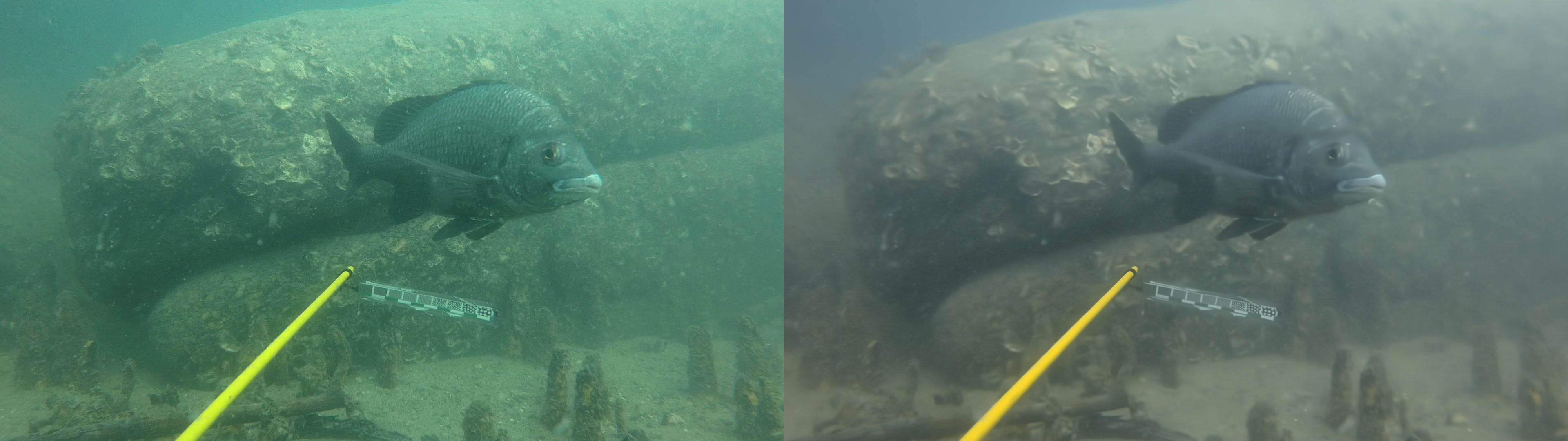}
\caption{Example  of the input and output of the trained IPT model.} \label{img}
\end{figure}

In this work, we utilize the pre-trained image transformer IPT \cite{ipt} for creating a model that performs image enhancement for underwater images based on the dataset UFO-120 \cite{ufo120}. The results show good performance (see an example in figure \ref{img}) indicating the potential of using a vision transformer for image enhancement. The paper is organized as follows. Section 2 discusses the related work for underwater image enhancement. In Section 3 we discuss the methodology of our work. Section 4 describes the experimental setup. Section 5 shows the results of this work and finally, Section 6  concludes the paper.


\section{Related work}
There have been many proposed works that performed image dehazing/enhancement using different approaches ranging from simple math-based technqiues to more complex systems using CNNs. One of the works that build up upon a basic yet effective concept is \cite{DCP} which utilizes the dark channel prior approach to estimate and remove the haze from an image. This method is based on the assumption that in an image with haze, some pixels (usually background pixels) will have a low-intensity value in at least one of the color channels that represent the haze in the image. By finding those pixels we can estimate the haze in the image and thus remove it considering it as an additive noise to the image. One drawback of this technique is that it depends on the estimation of the atmospheric light which might be invalid in some underwater lighting conditions. The work \cite{IBLA} also proposes a math-based approach but instead of relying on the dark channel prior, it utilizes the image blurriness map to estimate background light, transmission map, and depth estimation. Using those three estimations, an enhanced image can be produced from the underwater image. This method showed a noticeable improvement over the dark channel prior method. Since CNNs are continuously proving its robustness in many domains, some works have utilized it to perform underwater image enhancement. The work \cite{WaterNet} presents a "Water-Net" network that processes images at a deeper level. The network applies White balance, Histogram equalization, and Gamma Correction on the inputted image individually. The output of each  processes is then fed into a CNN network with 8 layers. The same output of the three processes will also be fed to three Feature Transformation Units (FTU) which have the purpose of decreasing the color casts and artifacts created by the these processes. The outputs of the CNN for each process will  be then combined with the corresponding output of the FTU to form the final enhanced image. This network showed the ability to adapt to different types of image distortion introduced by different underwater conditions. In a different approach,  \cite{U_GAN} proposed a Generative Adversarial Networks (GANs) which utilize the concept of game theory to create a network for enhancing underwater images. The network consists of two parts, the generator, and the discriminator. The generator will always work to create "fake" aiming to fool the discriminator while the discriminator will always try to differentiate between the "real" and "fake" data. In the case of underwater image enhancement, the generator will try to generate a copy of the haze image that looks similar to the clean image and the discriminator will try to distinguish between the clean image and the enhanced haze image thus it drives the generator to generate a further enhanced images that look as close to the clean images, and so on.  The network was tested on a diver tracking algorithm where it showed an improvement in the tracking performance  compared to when it is used with images without any enhancement. In a similar manner, \cite{FUnIE-GAN} proposed a GAN approach. The generator is based on the U-Net \cite{unet} and it consists of an encoder-decoder network with skip-connections between the mirrored layers  followed by Leaky-ReLU non-linearity and Batch Normalization. As for the discriminator, it was designed based on Markovian Patch-GAN architecture \cite{Markovian} which assumes the independence of pixels thus  improving the effectiveness of capturing high-frequency features in addition to requiring fewer parameters to compute. The proposed network showed higher performance for algorithms of object detection, saliency prediction, and human pose estimation. Last but not least,  \cite{ufo120} presents a CNN-based approach called Deep SESR that performs the best among the discussed methods on the UFO-120 dataset. The work proposes an end-to-end architecture that performs underwater image cleaning and up-sampling. The network mainly consists of Feature Extraction Network (FENet) block followed by an Auxiliary Attention Network (AAN) for saliency and convolutional layers for up-sampling. The FENet block aims to learn locally dense features while keeping shallow global architecture to guarantee fast feature extraction. The FENet consists of convolutional layers in addition to Residual Dense Blocks (RDBs) which in turn, consists of three sets of convolutional layers while the input and the output between each layer are concatenated. Such a design of the RDBs improves the ability to learn hierarchical features. The robust design of the Deep SESR network has achieved the highest PSNR and SSIM values on the UFO-120 dataset.   


\section{Methodology}
\begin{figure}
\includegraphics[width=\textwidth]{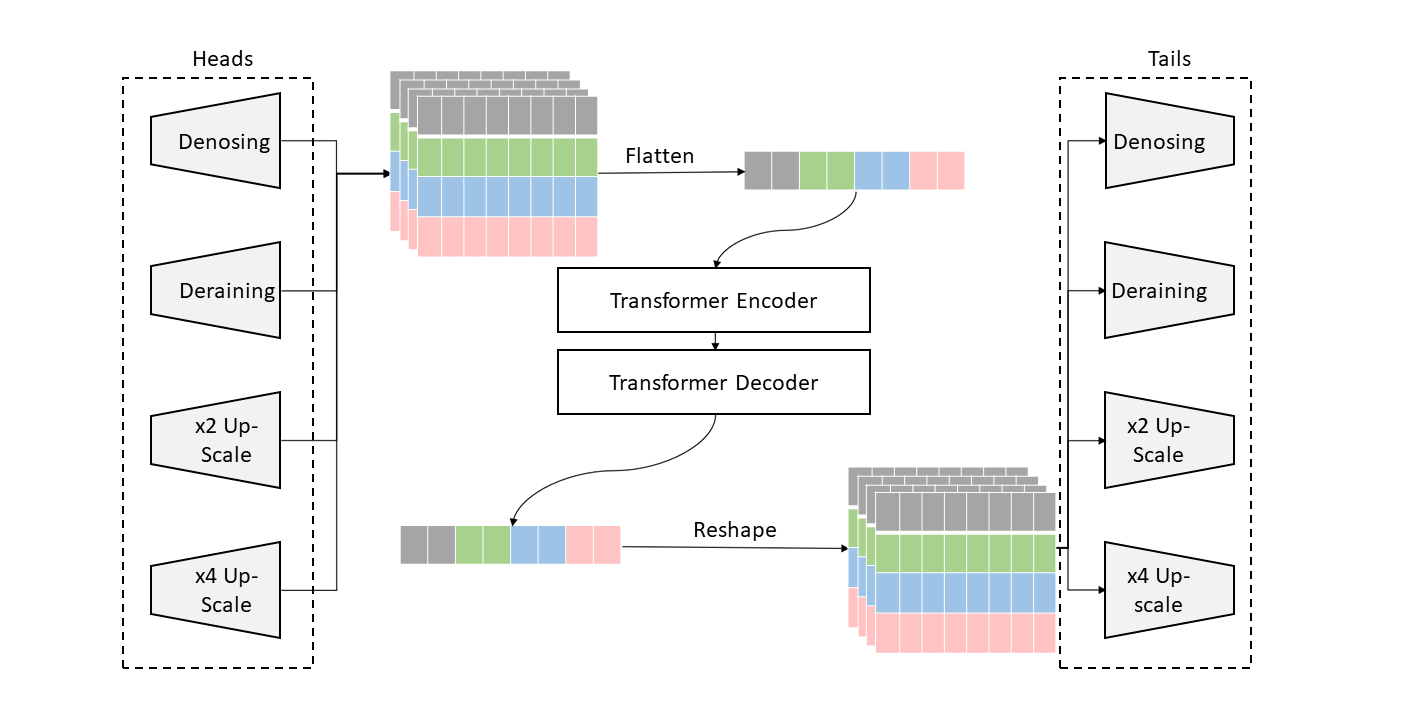}
\caption{The overall process of the proposed work \cite{ipt}.} \label{flow}
\end{figure}

The pre-trained image transformer \cite{ipt} showed  good results in denoising, deraining, and super-resolution that outperformed some state-of-art algorithms which was the reason for choosing this network. The IPT network has four main parts: heads, transformer encoder, transformer decoder, and tails. The first part contains four heads, each head consists of three convolutional layers and handles one task (denoising, deraining, *2 up-scale and *4 up-scale). Each head will generate a feature map and flatten it into patches before passing it to the transformer encoder. The transformer encoder has the same structure as in the work \cite{ipt2} which is based on a multi-head self-attention module with a feed-forward network. It follows that the transformer decoder that has the same structure as the encoder with the difference being the use of task-specific embedding as additional input. The output of the transformer decoder will be reshaped into the original input dimensions before passing to the next step. Last but not least, four tails for each of the four tasks similar to the heads. Patch normalization was not used. The general network can be seen in figure \ref{flow}.

The proposed network was trained on images from the ImageNet dataset. For each task, the images from the dataset were modified to train the network for each of the target tasks. For instance, to train the network for denoising, an image will be taken from the dataset regardless of its label. A Gaussian noise will be applied to the image, then the image with noise will be fed to the network as an input while the original image without the noise will be the ground truth. The same concept is applied to deraining where a "rain noise" is applied to the image. As for up-scaling, the inputted image is a down-scaled copy of the original image. The work presented a number of pre-trained models for two levels of noise, deraining, and super-resolution. The hardware used for creating the trained models consist of 32 NVIDIA Tesla V100 cards. 

Since the ground truth image is available during the training, a supervised fashion loss function was used as follow:
\begin{equation}
L_{supervised}=\sum_{i=1}^{N_{t}}L_{1}(IPT(I_{corrupted}^{i}),I_{clean})
\end{equation}
Where \(I_{clean}\) is the ground truth image and \(I_{corrupted}^{i}\) is the corrupted while \(i\) is one of the four tasks which indicate that the training of the different tasks occurs simultaneously. $L_1$ is the $L_1$ norm.
As for the optimizer, the Adam optimizer was used.

\begin{figure}
\centering
\begin{tikzpicture}
\begin{axis}[
    title={Loss during training},
    xlabel={Epoch},
    ylabel={Loss},
    xmin=0, xmax=5,
    ymin=0, ymax=20,
]

\addplot[
    color=blue,
    ]
    coordinates {(0.005,2.8176)(0.01,18.1786)(0.015,8.2534)(0.02,4.5324)(0.025,6.567)(0.03,3.9327)(0.035,1.7306)(0.04,2.8769)(0.045,2.4757)(0.05,1.4754)(0.055,1.7166)(0.06,1.3018)(0.065,1.1679)(0.07,1.4587)(0.075,0.7211)(0.08,1.4781)(0.085,1.1774)(0.09,0.6182)(0.095,0.9205)(0.1,0.9029)(0.105,0.797)(0.11,0.658)(0.115,0.5841)(0.12,0.6863)(0.125,0.5126)(0.13,0.5084)(0.135,0.6433)(0.14,0.4427)(0.145,0.5354)(0.15,0.4265)(0.155,0.4464)(0.16,0.4282)(0.165,0.4554)(0.17,0.3692)(0.175,0.3457)(0.18,0.3584)(0.185,0.3529)(0.19,0.2853)(0.195,0.3023)(0.2,0.2243)(0.205,0.2472)(0.21,0.2403)(0.215,0.2361)(0.22,0.2795)(0.225,0.2591)(0.23,0.2327)(0.235,0.1906)(0.24,0.2319)(0.245,0.2543)(0.25,0.255)(0.255,0.2856)(0.26,0.2695)(0.265,0.217)(0.27,0.2584)(0.275,0.2748)(0.28,0.2432)(0.285,0.2372)(0.29,0.2315)(0.295,0.2296)(0.3,0.2759)(0.305,0.2035)(0.31,0.3066)(0.315,0.2264)(0.32,0.2717)(0.325,0.2248)(0.33,0.2217)(0.335,0.2136)(0.34,0.1873)(0.345,0.2849)(0.35,0.3631)(0.355,0.2471)(0.36,0.2672)(0.365,0.2195)(0.37,0.2933)(0.375,0.2139)(0.38,0.225)(0.385,0.2492)(0.39,0.2983)(0.395,0.3247)(0.4,0.2568)(0.405,0.4295)(0.41,0.4311)(0.415,0.3977)(0.42,0.3365)(0.425,0.3787)(0.43,0.3321)(0.435,0.246)(0.44,0.2224)(0.445,0.3107)(0.45,0.2726)(0.455,0.2693)(0.46,0.2691)(0.465,0.2724)(0.47,0.2829)(0.475,0.2331)(0.48,0.2964)(0.485,0.2565)(0.49,0.2872)(0.495,0.307)(0.5,0.2702)(0.505,0.2719)(0.51,0.2987)(0.515,0.265)(0.52,0.3698)(0.525,0.2884)(0.53,0.4152)(0.535,0.3387)(0.54,0.422)(0.545,0.3113)(0.55,0.398)(0.555,0.2998)(0.56,0.4257)(0.565,0.3017)(0.57,0.4427)(0.575,0.4215)(0.58,0.5497)(0.585,0.2678)(0.59,0.5575)(0.595,0.5179)(0.6,0.4043)(0.605,0.4379)(0.61,0.3424)(0.615,0.4285)(0.62,0.3051)(0.625,0.2857)(0.63,0.319)(0.635,0.196)(0.64,0.2434)(0.645,0.2498)(0.65,0.1986)(0.655,0.2251)(0.66,0.2312)(0.665,0.185)(0.67,0.2104)(0.675,0.2187)(0.68,0.1734)(0.685,0.2002)(0.69,0.224)(0.695,0.2596)(0.7,0.1894)(0.705,0.2077)(0.71,0.2071)(0.715,0.1968)(0.72,0.2686)(0.725,0.1936)(0.73,0.2374)(0.735,0.1884)(0.74,0.1695)(0.745,0.1802)(0.75,0.1474)(0.755,0.161)(0.76,0.1345)(0.765,0.1547)(0.77,0.1093)(0.775,0.1442)(0.78,0.1305)(0.785,0.1411)(0.79,0.1743)(0.795,0.1989)(0.8,0.1679)(0.805,0.2424)(0.81,0.1758)(0.815,0.1482)(0.82,0.2625)(0.825,0.1835)(0.83,0.1372)(0.835,0.2662)(0.84,0.185)(0.845,0.1436)(0.85,0.3005)(0.855,0.172)(0.86,0.1766)(0.865,0.2506)(0.87,0.1672)(0.875,0.1497)(0.88,0.1864)(0.885,0.1667)(0.89,0.1357)(0.895,0.1333)(0.9,0.2043)(0.905,0.2424)(0.91,0.1395)(0.915,0.219)(0.92,0.238)(0.925,0.1729)(0.93,0.3287)(0.935,0.2161)(0.94,0.2877)(0.945,0.1568)(0.95,0.1747)(0.955,0.2464)(0.96,0.2193)(0.965,0.3113)(0.97,0.2207)(0.975,0.2726)(0.98,0.2427)(0.985,0.1865)(0.99,0.2981)(0.995,0.1487)(1,0.28)(1.005,0.1698)(1.01,0.2133)(1.015,0.1997)(1.02,0.1911)(1.025,0.4465)(1.03,0.325)(1.035,0.4249)(1.04,0.4403)(1.045,0.2826)(1.05,0.281)(1.055,0.3246)(1.06,0.2693)(1.065,0.3227)(1.07,0.2734)(1.075,0.3607)(1.08,0.2593)(1.085,0.3393)(1.09,0.2867)(1.095,0.3151)(1.1,0.36)(1.105,0.2255)(1.11,0.3611)(1.115,0.3806)(1.12,0.2816)(1.125,0.2204)(1.13,0.2504)(1.135,0.1965)(1.14,0.3144)(1.145,0.2673)(1.15,0.2003)(1.155,0.2741)(1.16,0.1432)(1.165,0.1857)(1.17,0.138)(1.175,0.1755)(1.18,0.1645)(1.185,0.1693)(1.19,0.2189)(1.195,0.188)(1.2,0.1456)(1.205,0.1227)(1.21,0.2524)(1.215,0.2888)(1.22,0.165)(1.225,0.264)(1.23,0.2156)(1.235,0.186)(1.24,0.3968)(1.245,0.2227)(1.25,0.2752)(1.255,0.2448)(1.26,0.2071)(1.265,0.2763)(1.27,0.0922)(1.275,0.2602)(1.28,0.2463)(1.285,0.2288)(1.29,0.313)(1.295,0.3147)(1.3,0.5371)(1.305,0.5346)(1.31,0.1843)(1.315,0.45)(1.32,0.2621)(1.325,0.4202)(1.33,0.3486)(1.335,0.2947)(1.34,0.3591)(1.345,0.2297)(1.35,0.2145)(1.355,0.1736)(1.36,0.2363)(1.365,0.2176)(1.37,0.2545)(1.375,0.158)(1.38,0.2225)(1.385,0.1875)(1.39,0.2053)(1.395,0.1843)(1.4,0.1871)(1.405,0.166)(1.41,0.1986)(1.415,0.2362)(1.42,0.1874)(1.425,0.2475)(1.43,0.1881)(1.435,0.2446)(1.44,0.1832)(1.445,0.2435)(1.45,0.153)(1.455,0.234)(1.46,0.2061)(1.465,0.2587)(1.47,0.1548)(1.475,0.2686)(1.48,0.1834)(1.485,0.2725)(1.49,0.1819)(1.495,0.2444)(1.5,0.1551)(1.505,0.2601)(1.51,0.1644)(1.515,0.2584)(1.52,0.1365)(1.525,0.2432)(1.53,0.144)(1.535,0.2179)(1.54,0.1491)(1.545,0.2703)(1.55,0.1614)(1.555,0.2319)(1.56,0.1602)(1.565,0.2158)(1.57,0.1272)(1.575,0.1304)(1.58,0.1447)(1.585,0.1275)(1.59,0.1413)(1.595,0.1232)(1.6,0.1734)(1.605,0.1278)(1.61,0.1602)(1.615,0.1366)(1.62,0.1747)(1.625,0.1428)(1.63,0.1775)(1.635,0.2132)(1.64,0.1453)(1.645,0.2805)(1.65,0.2032)(1.655,0.2744)(1.66,0.1893)(1.665,0.2829)(1.67,0.27)(1.675,0.2128)(1.68,0.2452)(1.685,0.1984)(1.69,0.2299)(1.695,0.1331)(1.7,0.1926)(1.705,0.1389)(1.71,0.1632)(1.715,0.1679)(1.72,0.1686)(1.725,0.1447)(1.73,0.1652)(1.735,0.1438)(1.74,0.1311)(1.745,0.1596)(1.75,0.1304)(1.755,0.1432)(1.76,0.1309)(1.765,0.1243)(1.77,0.1545)(1.775,0.1371)(1.78,0.1559)(1.785,0.1402)(1.79,0.1719)(1.795,0.2091)(1.8,0.2246)(1.805,0.2327)(1.81,0.1591)(1.815,0.2263)(1.82,0.2114)(1.825,0.1824)(1.83,0.1971)(1.835,0.2297)(1.84,0.1253)(1.845,0.2065)(1.85,0.1883)(1.855,0.1461)(1.86,0.1818)(1.865,0.1832)(1.87,0.1484)(1.875,0.1716)(1.88,0.1284)(1.885,0.1724)(1.89,0.177)(1.895,0.149)(1.9,0.1652)(1.905,0.1513)(1.91,0.1314)(1.915,0.1309)(1.92,0.1378)(1.925,0.1407)(1.93,0.1168)(1.935,0.1419)(1.94,0.1409)(1.945,0.132)(1.95,0.1316)(1.955,0.1412)(1.96,0.1341)(1.965,0.1553)(1.97,0.158)(1.975,0.1445)(1.98,0.1489)(1.985,0.1562)(1.99,0.1017)(1.995,0.1397)(2,0.1161)(2.005,0.1204)(2.01,0.126)(2.015,0.1089)(2.02,0.3606)(2.025,0.2233)(2.03,0.1879)(2.035,0.1691)(2.04,0.2483)(2.045,0.1326)(2.05,0.2138)(2.055,0.2229)(2.06,0.2031)(2.065,0.1513)(2.07,0.2488)(2.075,0.2167)(2.08,0.146)(2.085,0.2894)(2.09,0.241)(2.095,0.2234)(2.1,0.2431)(2.105,0.1805)(2.11,0.1869)(2.115,0.1898)(2.12,0.1767)(2.125,0.2723)(2.13,0.1406)(2.135,0.248)(2.14,0.1514)(2.145,0.2924)(2.15,0.242)(2.155,0.2693)(2.16,0.3036)(2.165,0.1256)(2.17,0.1807)(2.175,0.1783)(2.18,0.118)(2.185,0.2162)(2.19,0.1872)(2.195,0.1247)(2.2,0.1771)(2.205,0.1673)(2.21,0.1446)(2.215,0.1711)(2.22,0.1561)(2.225,0.1261)(2.23,0.1607)(2.235,0.1553)(2.24,0.2156)(2.245,0.1482)(2.25,0.1606)(2.255,0.138)(2.26,0.1502)(2.265,0.1353)(2.27,0.1293)(2.275,0.1318)(2.28,0.1184)(2.285,0.1462)(2.29,0.0948)(2.295,0.173)(2.3,0.2024)(2.305,0.1322)(2.31,0.2279)(2.315,0.1564)(2.32,0.2537)(2.325,0.2261)(2.33,0.1938)(2.335,0.2301)(2.34,0.1467)(2.345,0.1794)(2.35,0.164)(2.355,0.0996)(2.36,0.1243)(2.365,0.1241)(2.37,0.0953)(2.375,0.1339)(2.38,0.1589)(2.385,0.1587)(2.39,0.1438)(2.395,0.1463)(2.4,0.1295)(2.405,0.124)(2.41,0.2297)(2.415,0.1549)(2.42,0.2007)(2.425,0.211)(2.43,0.1581)(2.435,0.282)(2.44,0.1556)(2.445,0.2782)(2.45,0.1656)(2.455,0.2755)(2.46,0.1809)(2.465,0.1403)(2.47,0.2163)(2.475,0.1939)(2.48,0.1694)(2.485,0.2038)(2.49,0.1274)(2.495,0.1462)(2.5,0.131)(2.505,0.1407)(2.51,0.1277)(2.515,0.1257)(2.52,0.1167)(2.525,0.141)(2.53,0.127)(2.535,0.121)(2.54,0.1272)(2.545,0.148)(2.55,0.151)(2.555,0.1149)(2.56,0.1685)(2.565,0.1176)(2.57,0.1481)(2.575,0.1513)(2.58,0.1282)(2.585,0.1379)(2.59,0.1564)(2.595,0.1201)(2.6,0.1792)(2.605,0.1468)(2.61,0.1576)(2.615,0.1443)(2.62,0.1537)(2.625,0.1053)(2.63,0.089)(2.635,0.0874)(2.64,0.0633)(2.645,0.0773)(2.65,0.0836)(2.655,0.1166)(2.66,0.1143)(2.665,0.1266)(2.67,0.1144)(2.675,0.1209)(2.68,0.1138)(2.685,0.1414)(2.69,0.1549)(2.695,0.1518)(2.7,0.146)(2.705,0.1457)(2.71,0.1084)(2.715,0.1688)(2.72,0.1121)(2.725,0.1886)(2.73,0.1158)(2.735,0.2061)(2.74,0.1585)(2.745,0.1794)(2.75,0.1253)(2.755,0.2125)(2.76,0.1635)(2.765,0.2732)(2.77,0.2588)(2.775,0.1581)(2.78,0.2245)(2.785,0.1628)(2.79,0.1488)(2.795,0.1674)(2.8,0.1694)(2.805,0.1528)(2.81,0.1312)(2.815,0.1683)(2.82,0.1278)(2.825,0.1086)(2.83,0.1317)(2.835,0.12)(2.84,0.1013)(2.845,0.1229)(2.85,0.1416)(2.855,0.159)(2.86,0.1763)(2.865,0.1428)(2.87,0.1265)(2.875,0.153)(2.88,0.1098)(2.885,0.1383)(2.89,0.1422)(2.895,0.0941)(2.9,0.1225)(2.905,0.1421)(2.91,0.2126)(2.915,0.222)(2.92,0.136)(2.925,0.1697)(2.93,0.1905)(2.935,0.1368)(2.94,0.1675)(2.945,0.1941)(2.95,0.1277)(2.955,0.1588)(2.96,0.1691)(2.965,0.1733)(2.97,0.1793)(2.975,0.1464)(2.98,0.1519)(2.985,0.1814)(2.99,0.1193)(2.995,0.1442)(3,0.1351)(3.005,0.1299)(3.01,0.146)(3.015,0.1562)(3.02,0.2474)(3.025,0.2045)(3.03,0.1958)(3.035,0.1803)(3.04,0.2168)(3.045,0.1558)(3.05,0.1479)(3.055,0.2162)(3.06,0.1634)(3.065,0.1141)(3.07,0.1556)(3.075,0.156)(3.08,0.1457)(3.085,0.1481)(3.09,0.1437)(3.095,0.1289)(3.1,0.1313)(3.105,0.1088)(3.11,0.1496)(3.115,0.1115)(3.12,0.1376)(3.125,0.1231)(3.13,0.1649)(3.135,0.1695)(3.14,0.1378)(3.145,0.1523)(3.15,0.144)(3.155,0.1205)(3.16,0.1116)(3.165,0.1102)(3.17,0.1101)(3.175,0.0885)(3.18,0.1503)(3.185,0.1104)(3.19,0.1)(3.195,0.1037)(3.2,0.0996)(3.205,0.0965)(3.21,0.1212)(3.215,0.0932)(3.22,0.1118)(3.225,0.1042)(3.23,0.0951)(3.235,0.109)(3.24,0.1485)(3.245,0.1421)(3.25,0.1143)(3.255,0.1404)(3.26,0.1228)(3.265,0.1561)(3.27,0.1498)(3.275,0.1238)(3.28,0.1661)(3.285,0.1381)(3.29,0.1114)(3.295,0.1724)(3.3,0.2036)(3.305,0.1364)(3.31,0.1109)(3.315,0.1762)(3.32,0.1295)(3.325,0.1178)(3.33,0.1648)(3.335,0.0876)(3.34,0.1326)(3.345,0.1945)(3.35,0.1202)(3.355,0.1263)(3.36,0.138)(3.365,0.1267)(3.37,0.1063)(3.375,0.1239)(3.38,0.0664)(3.385,0.0724)(3.39,0.0846)(3.395,0.0643)(3.4,0.0793)(3.405,0.1508)(3.41,0.1361)(3.415,0.1783)(3.42,0.1194)(3.425,0.1729)(3.43,0.1265)(3.435,0.1777)(3.44,0.1059)(3.445,0.1783)(3.45,0.1308)(3.455,0.1318)(3.46,0.0949)(3.465,0.1413)(3.47,0.1136)(3.475,0.1177)(3.48,0.1271)(3.485,0.1087)(3.49,0.1159)(3.495,0.0969)(3.5,0.1095)(3.505,0.096)(3.51,0.1083)(3.515,0.1062)(3.52,0.1227)(3.525,0.1195)(3.53,0.121)(3.535,0.1075)(3.54,0.1079)(3.545,0.0903)(3.55,0.1064)(3.555,0.087)(3.56,0.1328)(3.565,0.0871)(3.57,0.1428)(3.575,0.1098)(3.58,0.1118)(3.585,0.1099)(3.59,0.1114)(3.595,0.1083)(3.6,0.0982)(3.605,0.0969)(3.61,0.1014)(3.615,0.0923)(3.62,0.0957)(3.625,0.0895)(3.63,0.1253)(3.635,0.1246)(3.64,0.1195)(3.645,0.1273)(3.65,0.1193)(3.655,0.0882)(3.66,0.0784)(3.665,0.0908)(3.67,0.08)(3.675,0.0889)(3.68,0.0981)(3.685,0.0875)(3.69,0.1406)(3.695,0.0729)(3.7,0.162)(3.705,0.092)(3.71,0.2789)(3.715,0.1353)(3.72,0.2212)(3.725,0.1668)(3.73,0.2026)(3.735,0.1574)(3.74,0.0973)(3.745,0.1447)(3.75,0.1134)(3.755,0.1312)(3.76,0.1154)(3.765,0.1194)(3.77,0.1183)(3.775,0.1259)(3.78,0.1081)(3.785,0.1244)(3.79,0.1062)(3.795,0.1499)(3.8,0.1161)(3.805,0.1351)(3.81,0.1113)(3.815,0.1239)(3.82,0.1532)(3.825,0.1021)(3.83,0.1488)(3.835,0.1042)(3.84,0.1442)(3.845,0.1034)(3.85,0.1697)(3.855,0.0959)(3.86,0.1645)(3.865,0.1171)(3.87,0.1585)(3.875,0.1183)(3.88,0.139)(3.885,0.1154)(3.89,0.1208)(3.895,0.1015)(3.9,0.1063)(3.905,0.1034)(3.91,0.1099)(3.915,0.1126)(3.92,0.1042)(3.925,0.1078)(3.93,0.1042)(3.935,0.0765)(3.94,0.1091)(3.945,0.0804)(3.95,0.1026)(3.955,0.0946)(3.96,0.1303)(3.965,0.1206)(3.97,0.1372)(3.975,0.1188)(3.98,0.1304)(3.985,0.1097)(3.99,0.1218)(3.995,0.09)(4,0.0997)(4.005,0.1004)(4.01,0.0814)(4.015,0.1173)(4.02,0.1264)(4.025,0.0932)(4.03,0.1184)(4.035,0.1119)(4.04,0.096)(4.045,0.1087)(4.05,0.0895)(4.055,0.1032)(4.06,0.0897)(4.065,0.093)(4.07,0.1346)(4.075,0.1471)(4.08,0.1094)(4.085,0.1397)(4.09,0.1254)(4.095,0.0999)(4.1,0.1128)(4.105,0.1102)(4.11,0.1073)(4.115,0.0997)(4.12,0.1018)(4.125,0.0953)(4.13,0.1622)(4.135,0.1401)(4.14,0.1391)(4.145,0.1309)(4.15,0.1274)(4.155,0.0857)(4.16,0.1054)(4.165,0.0978)(4.17,0.0879)(4.175,0.1206)(4.18,0.1169)(4.185,0.0871)(4.19,0.1187)(4.195,0.0917)(4.2,0.1155)(4.205,0.092)(4.21,0.1114)(4.215,0.0795)(4.22,0.1092)(4.225,0.1091)(4.23,0.0971)(4.235,0.1028)(4.24,0.1098)(4.245,0.0968)(4.25,0.1048)(4.255,0.0836)(4.26,0.1014)(4.265,0.1657)(4.27,0.1283)(4.275,0.1623)(4.28,0.1084)(4.285,0.1431)(4.29,0.1045)(4.295,0.0776)(4.3,0.0698)(4.305,0.0789)(4.31,0.0729)(4.315,0.0823)(4.32,0.0861)(4.325,0.1052)(4.33,0.0852)(4.335,0.1102)(4.34,0.1045)(4.345,0.1053)(4.35,0.0835)(4.355,0.1046)(4.36,0.0802)(4.365,0.1147)(4.37,0.1054)(4.375,0.1737)(4.38,0.1157)(4.385,0.1744)(4.39,0.0964)(4.395,0.1533)(4.4,0.1398)(4.405,0.1309)(4.41,0.1075)(4.415,0.1213)(4.42,0.0986)(4.425,0.1012)(4.43,0.0991)(4.435,0.101)(4.44,0.1078)(4.445,0.0906)(4.45,0.1001)(4.455,0.0996)(4.46,0.1274)(4.465,0.1724)(4.47,0.1172)(4.475,0.1577)(4.48,0.1152)(4.485,0.1594)(4.49,0.0719)(4.495,0.1007)(4.5,0.0869)(4.505,0.0959)(4.51,0.0868)(4.515,0.0889)(4.52,0.1914)(4.525,0.1631)(4.53,0.1488)(4.535,0.155)(4.54,0.1637)(4.545,0.0973)(4.55,0.1077)(4.555,0.1232)(4.56,0.0871)(4.565,0.1178)(4.57,0.1332)(4.575,0.1394)(4.58,0.1593)(4.585,0.128)(4.59,0.1166)(4.595,0.1395)(4.6,0.1603)(4.605,0.1516)(4.61,0.1613)(4.615,0.1434)(4.62,0.1412)(4.625,0.1237)(4.63,0.1036)(4.635,0.1175)(4.64,0.1212)(4.645,0.0902)(4.65,0.1038)(4.655,0.092)(4.66,0.1087)(4.665,0.0997)(4.67,0.0885)(4.675,0.1221)(4.68,0.129)(4.685,0.2042)(4.69,0.1663)(4.695,0.1506)(4.7,0.1757)(4.705,0.1113)(4.71,0.1484)(4.715,0.0957)(4.72,0.1115)(4.725,0.1017)(4.73,0.0847)(4.735,0.1229)(4.74,0.087)(4.745,0.0852)(4.75,0.0896)(4.755,0.0798)(4.76,0.0865)(4.765,0.1254)(4.77,0.1205)(4.775,0.1291)(4.78,0.12)(4.785,0.1258)(4.79,0.0905)(4.795,0.1123)(4.8,0.0911)(4.805,0.1006)(4.81,0.0988)(4.815,0.0918)(4.82,0.0927)(4.825,0.0983)(4.83,0.094)(4.835,0.0939)(4.84,0.0951)(4.845,0.1094)(4.85,0.1151)(4.855,0.0825)(4.86,0.1173)(4.865,0.0815)(4.87,0.0929)(4.875,0.113)(4.88,0.0849)(4.885,0.096)(4.89,0.0979)(4.895,0.0868)(4.9,0.1385)(4.905,0.1056)(4.91,0.1421)(4.915,0.1146)(4.92,0.1383)(4.925,0.0945)(4.93,0.1986)(4.935,0.1648)(4.94,0.1529)(4.945,0.1852)(4.95,0.1241)(4.955,0.1386)(4.96,0.0908)(4.965,0.0786)(4.97,0.0825)(4.975,0.0828)(4.98,0.0781)(4.985,0.0892)(4.99,0.0842)(4.995,0.0808)(5,0.0876)};

\end{axis}
\end{tikzpicture}
\caption{The loss values during the training process of 5 epochs.} \label{loss}
\end{figure}
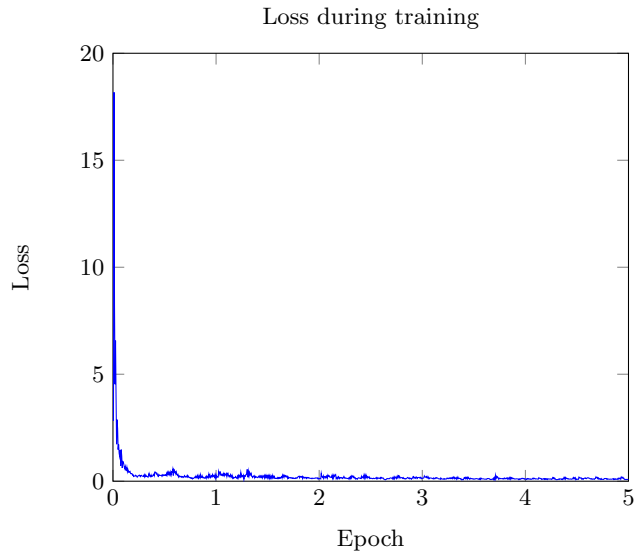

\section{Experiment}
The pre-trained transformer came with pre-trained models for the four previously mention tasks (denosing, deraining, 2x up-scaling and 4x up-scaling). Since we have limited processing power, we will fine-tune one of the pre-trained models. The model that gave the best initial results was the model for denosing with Gaussian noise of (\(\Sigma=30\)) which make sense since the distortion in underwater images is basically a noise in the image, we fine tune the pre-trained model with images from underwater dataset. The dataset that will be used is UFO-10 \cite{ufo120}. This dataset was created for three main tasks: image enhancement, super-resolution and object segmentation. The dataset consists of 1500 pairs of clear and non-clear underwater images as well as a mask for the objects in the image. The clear images come with higher resolution in comparison to the non-clear images for the up-scaling task. Since we do not attempt to train for up-scaling, we applied a smoothing filter with a size of 
3 $\times$ 3 so both clear and non-clear images have the same level of details. In addition to that, data augmentation was applied in the form of rotation of 5 different angles (0, 45, 135, 225 and 315) to make the total number of images pairs (haze \& ground truth) 7500. 6000 images pairs were taken for training while the rest 1500 images pairs were reserved for testing.

\begin{figure*}[ht]
\centering
\includegraphics[width=1\linewidth]{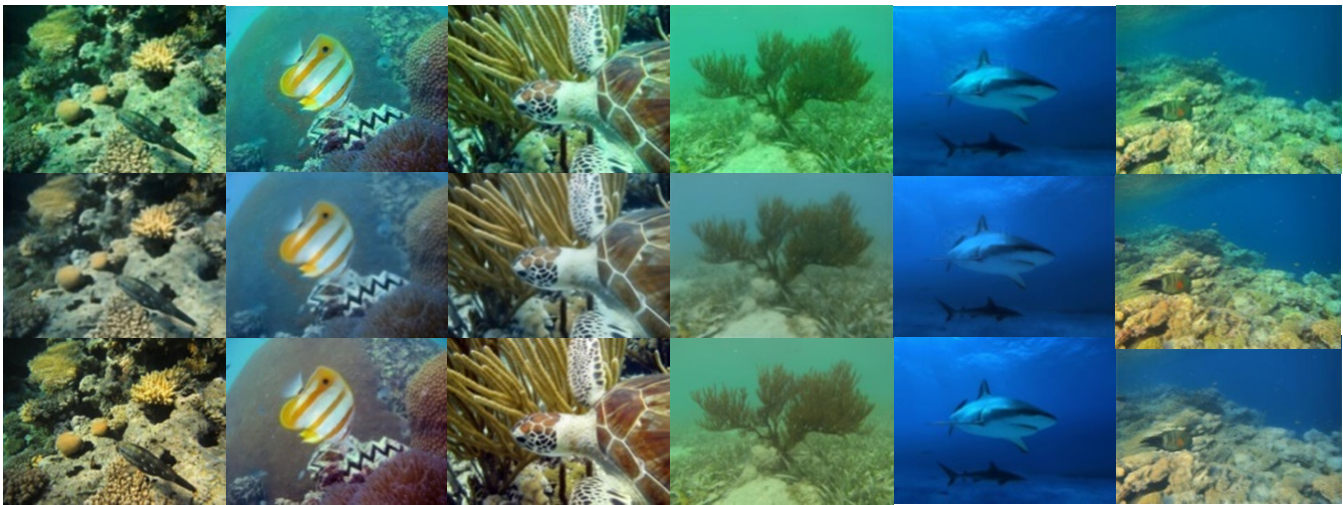}
\caption{Some of the results on the testing images where the top row represents the input image, the middle row represents the output image and the bottom row represents the ground truth.}
\label{results}
\end{figure*}

\section{Results}
We fine tuned the pre-trained model on the augmented underwater dataset resulted using NVIDIA Quadro P5000 GPU with 16Gb of graphic memory and 80 Gb RAM. After testing with many training configurations we found that training for 6 epochs with a batch size of one gave the best result. The resulting loss graph can be seen in figure \ref{loss}.

\begin{table}
\caption{The Peak Signal to Noise Ratio (PSNR) and Structural Similarity Index Measurement (SSIM) values for different methods on the UFO-120\cite{ufo120} dataset. The highest values is marked with bold.}\label{tab1}
\begin{tabularx}{\textwidth} { 
  | >{\centering\arraybackslash}X 
  | >{\centering\arraybackslash}X 
  | >{\centering\arraybackslash}X | }
 \hline
  Method &  PSNR & SSIM\\
\hline
DCP \cite{DCP}             & 18.20 & 0.71  \\ 
IBLA \cite{IBLA}           & 17.50  & 0.65  \\ 
WaterNet \cite{WaterNet}   & 22.46  & 0.79 \\ 
U-GAN \cite{U_GAN}         & 23.45  & 0.80 \\ 
FUnIE-GAN \cite{FUnIE-GAN} & 25.15  & 0.82  \\ 
Deep SESR \cite{ufo120}    & \textbf{27.15}  & 0.84 \\ 
Our approach                       & 23.14  &\textbf{0.90} \\ 
\hline
\end{tabularx}
\end{table}

We report some qualitative results in figure \ref{results}.
The top row represents the input images, the middle row represents the output images and the bottom row represents the ground truth. We noticed that the trained model works best with images that contain green tint where it removes most of the color distortion. In addition to that, we have noticed that the output images have slightly less details compared  to the input images. This slight loss of details is due  to the smoothing effect of the base model that was used for training. The pre-trained transformer network came with up-scaling model which could be used to restore some of the details in the image. However, that will almost double the processing time which reduce the ability of implementing it in a real-time system. On average, processing one image (320*240) took around 12 to 13 seconds which is not practical for real-time processing. In a future work, we will try different approached to reduce the processing time to for real-time applications.

For a quantitative  comparison with other  approaches for underwater imaging enhancing, we tested our algorithm  on UFO-120 dataset. We selected two measurements: Peak Signal to Noise Ratio (PSNR) and Structural Similarity Index Measurement (SSIM). Table \ref{tab1} shows the PSNR and SSIM values for the most relevant approaches found in the literature compared  to our work. 

It is evident that our approach outperforms the others in terms of SSIM, and came in fourth place in terms of PSNR. We believe that the difference if performance between SSIM and PSNR is due the fact that the trained model perform well in restoring the structure and color of the image. However, the smoothing effect reduced the details on the image hence the relatively low PSNR value.

\section{Conclusions}
Underwater  operations, such as exploration, monitoring and recovery, performed by autonomous or semiautonomos robots strongly rely on computer vision. However, the underwater environment is specifically challenging, due to poor visibility induced by relevant light absorption and scattering caused by  suspended particles that make the images blurry, bluish and with wide opaque background. For all these reasons, image enhancement of underwater images play a critical role for any mission of marine robotics. Many techniques have been proposed in the literature, mostly based on convolutional neural networks. 
In this work, we propose a different solution for image enhancement of underwater images, based on the image transformer network (IPT) \cite{ipt}. 
The underwater imaging dataset UFO-120 \cite{ufo120} has been selected for this study. A process of data augmentation allowed to increase the original size of 1500 i
mages to 7500 images. The augmented dataset has been used for training the transformer. 

We evaluated our results using two measurements (peak-signal-to-noise ratio (PSNR) and structure similarity index measurement (SSIM). This allowed us to compare the results with the same indexes reported in similar works using different approaches. We find that our approach outperforms the previous methodologies proposed in the literature in terms of SSIM and it is fourth in terms of PSNR. 

In the future work we will test our approach on different datasets from the underwater environments and we will also evaluate the effects on specific operative tasks, like object recognition and object tracking, when applying these technologies in an operative scenario, based on the use of semi-autonomous/autonomous robots.


\section*{Acknowledgment}
This work acknowledges the support provided by the Khalifa University of Science and Technology under awards No. RC1-2018-KUCARS, and CIRA-2019-047. The second author, Sajid Javed, of this publication is supported by the FSU -2022-003 Project under Award No. 000628-00001.

%
%

\bibliographystyle{splncs04}
\bibliography{REFERENCES}

\end{document}